\documentclass[11pt]{article}
\usepackage{moriond}
\usepackage{color}
\usepackage{graphicx}

\def\Journal#1#2#3#4{{#1} {\bf #2}, #3 (#4)}

\def\NIM{\em Nucl. Instrum. Methods}

\def\NP{{\em Nucl. Phys.}}

\def\be{\begin{equation}}
\def\ee{\end{equation}}
\def\bea{\begin{eqnarray}}
\def\eea{\end{eqnarray}}

\begin{document}

\vspace*{4cm}
\title{STRANGE PARTICLE PRODUCTION IN NUCLEAR COLLISIONS AT CERN-NA49}

\author{D. BARNA \emph{for the NA49 Collaboration} }

\address{Central Research Institute for Physics, Konkoly-Thege \'ut
  29-33, H-1121, Budapest, Hungary}

\maketitle\abstracts{  In order  to  quantify isospin  effects in  the
comparison  of   elementary  to   nuclear  collisions,  p-p   and  n-p
interactions have been  studied in the NA49 experiment  in addition to
p-Pb and Pb-Pb reactions.  Together with first measurements of cascade
hyperon and  $\Omega$ production in  p-p collisions, isospin-corrected
$K/\pi$    ratios   and   cascade    yields   are    discussed.    The
$\overline{\Omega}^+/\Omega^-$ ratio in p-p  collisions is found to be
less than 0.5 with 95\% confidence level.  }

\section{Introduction}

NA49    is    a    general    purpose    large    acceptance    hadron
spectrometer\cite{nim}. It's main tracking  devices are 4 large volume
TPCs, which make particle  identification possible. The experiment has
collected data  on a  wide range of  hadronic interactions  (p-p, d-p,
$\pi$-p,  p-Pb, $\pi$-Pb,  Pb-Pb).   These  data are  used  to make  a
systematic study and understand the complex A-A interactions.

\section{Isospin Effects on the $K$ and $\pi$ Production}

\begin{figure}[b]
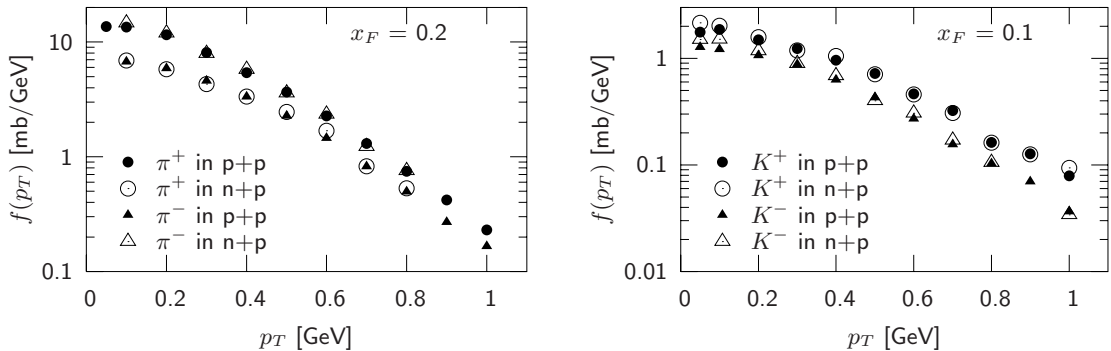

 \centerline{
   \footnotesize\sf
   \input{pion-in-np-pp}
   \input{kaon-in-np-pp}
 }
\caption{Pion (left) and kaon (right) invariant $p_T$
   distributions in p-p and n-p collisions (preliminary)} 
\label{fig:kaon-pion-in-pp-np}
\end{figure}

In order to trace the  signatures of new physics, heavy ion collisions
are usually compared  to elementary (p-p or p-A)  collisions. Beyond the
possible  'interesting' differences  there is  a  'trivial' difference
however: heavy ions  are composed of neutrons and  protons (for Pb the
mixture  is 60\% n,  40\% p).  In order  to study  differences between
proton and neutron fragmentation, NA49 has collected data with a deuteron
beam      incident      upon       a      proton      target      with
$p_{\mbox{\scriptsize beam}}$=158~GeV/nucleon. By tagging on  a spectator proton with
$p \approx$~158~GeV one can select neutron-proton collisions.

\begin{figure}
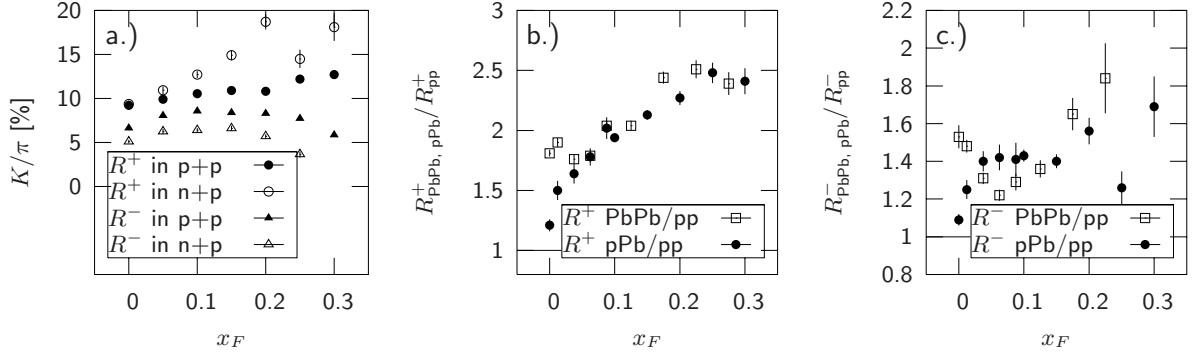

  \centerline{
    \footnotesize\sf
    \input{k-per-pi}
    \input{kplus-per-piplus-Pb}
    \input{kminus-per-piminus-Pb}
  }
  \caption{
    (a): the $R^+=K^+/\pi^+$ and $R^-=K^-/\pi^-$ ratio in pp and np
    collisions. (b) and (c): the $R^+$ and $R^-$ ratio in pPb and PbPb
    collisions, compared to the same ratio in pp collisions. The
    estimated isospin-effect was removed from the PbPb results, see
    text for details (preliminary results)}
  \label{fig:k-per-pi}
\end{figure}

Charged kaon  and pion yields were determined  via dE/dx measurements.
Figure  \ref{fig:kaon-pion-in-pp-np}  shows  the  transverse  momentum
distributions  in  p-p and  n-p collisions  in the
projectile  (n  or p)  hemisphere  at  fixed  Feynman-x. The  flip  of
positive and negative pions and the unchanged kaon yields, when  replacing the proton projectile by a
neutron projectile, are clearly visible.  Consequently
the $R^+ = K^+/\pi^+$ ratio is higher, the $R^- = K^-/\pi^-$ ratio is lower in n
than in p fragmentation (fig. \ref{fig:k-per-pi}a). 
A properly weighted  average  (40\%p + 60\%n fragmentation)
is  used then to  estimate the  magnitude of  isospin effects  in Pb-Pb
collisions.   Figures  \ref{fig:k-per-pi}  b.)   and   c.)  show   the
$R^\pm_{\mbox{\scriptsize pPb, PbPb}} / R^\pm_{\mbox{\scriptsize pp}}$
double ratios as a function of Feynman-x,  after the elimination  of the isospin effect  from the
Pb-Pb results. Note that
above $x_F=0.05$ the p-Pb and Pb-Pb results agree within errors. 
The deviation close to $x_F=0$ is due to the target feed-over: target
nucleons in p-Pb and Pb-Pb behave differently. They suffer 1 (in p-Pb)
or several  collisions (in Pb-Pb). 

\section{Strange Baryon Production}

\subsection{Strangeness Enhancement}

\begin{figure}
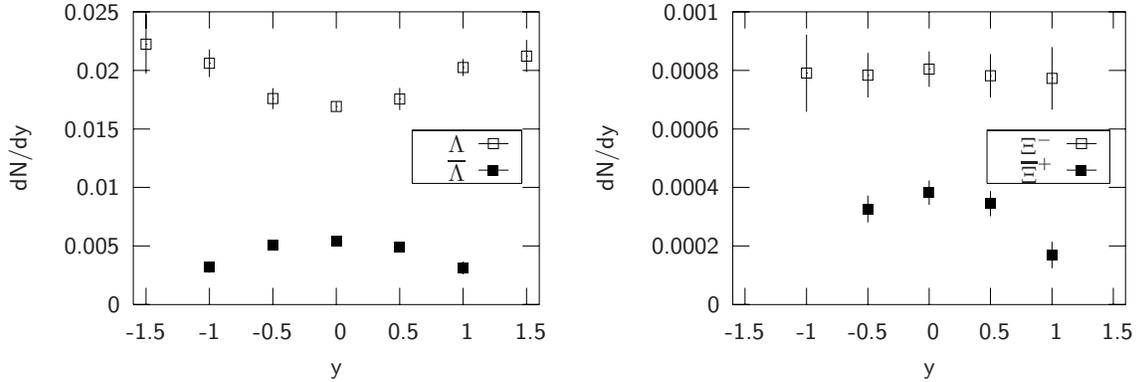

  \centerline{
    \footnotesize\sf
    \input{lambda-pp}\hss\input{xi-pp}
  }
  \caption{Rapidity distributions of $\Lambda$, $\overline{\Lambda}$,
    $\Xi$ and $\overline{\Xi}$ particles in p+p collisions}
  \label{fig:strange-baryons-pp}
\end{figure}

Strange and multistrange baryons are reconstructed via their weak
decay, using two independent reconstruction methods. The acceptance
extends over a wide rapidity range ($\pm$1 unit around midrapidity),
and covers the full $p_T$ range. Rapidity distributions of $\Lambda$, $\overline{\Lambda}$, $\Xi$ and
$\overline{\Xi}$ are shown in figure \ref{fig:strange-baryons-pp}. The $\Lambda$ ($\overline{\Lambda}$)
results have been feeddown-corrected with respect to the
$\Xi^-, \Xi^0\to\Lambda$ ($\overline{\Xi}^+,\overline{\Xi}^0\to\overline{\Lambda}$)
channel. For this the measured $\Xi^-$ and $\overline{\Xi}^+$ yields
were used. The $\Xi^0$ ($\overline{\Xi}^0$) yields were
estimated as $\Xi^0 = 1.5\cdot\Xi^-$ and $\overline{\Xi}^0 =
0.66\cdot\overline{\Xi}^+$. 

Strangeness enhancement  in heavy ion  collisions is often  defined as
the increase of  the strangeness/number-of-participants ratio compared
to p-p  collisions. Strangeness enhancement  can be attributed  to the
presence of  the Quark Gluon Plasma, or  the higher number  of collisions suffered  by an
average participant. Assuming the latter, p-A data can be used to make
a  study of  enhancement as  a function  of the  number  of collisions
$\nu$.

\begin{figure}
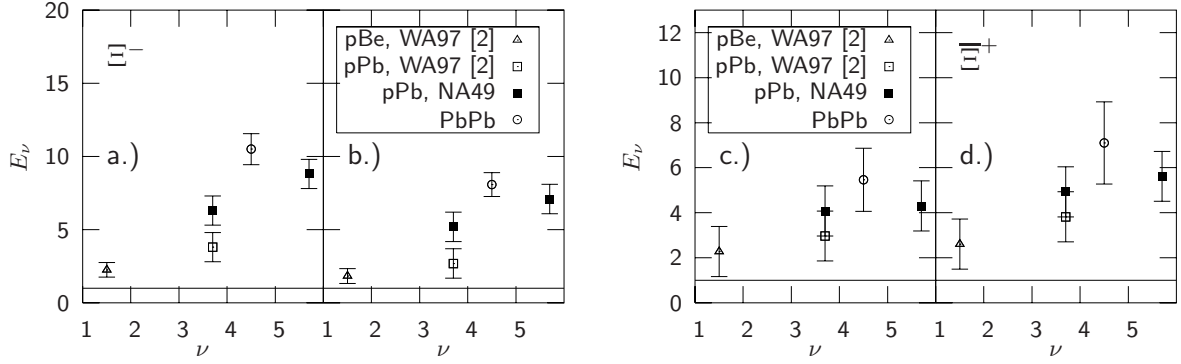

  \centerline{
    \footnotesize\sf
    {\input{enhancement-xi}} \hfill
    {\input{enhancement-axi}}
  }
  \caption{$\Xi^-$ (left panel) and $\overline{\Xi}^+$ (right panel)
    enhancement: (a) and (c) -- without isospin correction, (b) and (d)
    -- with isospin correction. The Pb-Pb point is an average of
    WA97\cite{wa97} and NA49\cite{na49.strange} results.}
  \label{fig:enhancement}
\end{figure}

Midrapidity  yields are  contributed to  by  both the  target
($T$) and  the
projectile ($P$).  In  the  symmetric  p-p  collisions  both contributions  are  therefore
$T_{pp} = P_{pp} = \frac{1}{2}\cdot\left.\frac{dN}{dy}\right|_{pp}^{y=0}$.   In the asymmetric
p-A collisions  no enhancement  per nucleon is  assumed on  the target
side (since each target nucleon  suffers only 1 collision). The target
contribution ($T_{pA}$) is  therefore  assumed  to be  the  target  contribution
derived  from the p-p  results ($T_{pp}$), scaled  by  the
number of  colliding target nuclei ($\nu$).  All the enhancement  ($E_\nu$) is
attributed to the projectile, suffering $\nu$ collisions:
$\left.\frac{dN}{dy}\right|_{pA}^{y=0} = \nu \cdot \frac{1}{2}
\left.\frac{dN}{dy}\right|_{pp}^{y=0} + E_\nu \cdot \frac{1}{2}
\left.\frac{dN}{dy}\right|_{pp}^{y=0}$.
The enhancement in the symmetric A-A collisions is defined as
$\left.\frac{dN}{dy}\right|_{AA}^{y=0} = E_\nu \cdot  \frac{N_W}{2} \cdot
\left.\frac{dN}{dy}\right|_{pp}^{y=0}$, where $N_W$ is the number of
wounded nucleons. Panels (a) and (c) of figure \ref{fig:enhancement} show
the enhancement $E_\nu$ for $\Xi^-$ and $\overline{\Xi}^+$  in p-Be, p-Pb and Pb-Pb
collisions as a
function of $\nu$ (which is the average number of collisions suffered
by the
projectile for p-A collisions, and by
an average nucleon for A-A collisions). Enhancement factors for $\Xi^-$ are larger than for
$\overline{\Xi}^+$. The enhancement factors for both the baryon and
the antibaryon are of the same order in p-Pb and Pb-Pb collisions. 

Since $\Xi$ production from a neutron or a proton can be different,
isospin effects can also play a role in strangeness enhancement. 
In order to take this into account, we redefined strangeness
enhancement, with this isospin effect detached:

\begin{eqnarray}
\left.\frac{dN}{dy}\right|_{pA}^{y=0} & = &
  \nu\cdot\alpha\cdot\frac{1}{2}\cdot\left.\frac{dN}{dy}\right|_{pp}^{y=0}
  + E_\nu\cdot\frac{1}{2} \cdot \left.\frac{dN}{dy}\right|_{pp}^{y=0}
  \label{eq:enhancement-pA-iso}
  \\
\left.\frac{dN}{dy}\right|_{AA}^{y=0} & = &
  E_\nu\cdot\frac{N_W}{2} \cdot \alpha \cdot \left.\frac{dN}{dy}\right|_{pp}^{y=0}
  \label{eq:enhancement-AA-iso}
\end{eqnarray}

\noindent where the $\alpha$ factor accounts for the increase or
  decrease of the $\Xi$ yield due to the proton-neutron composition of
  the lead nucleus. This could be derived by comparing an appropriately
  weighted average of $\Xi$ yields from proton (40\%) and neutron
  (60\%) induced reactions to the $\Xi$ yield in proton-induced
  reactions. Our limited statistics of n+p data did not allow a measurement of
  $\Xi$ production, therefore the $\alpha$ factor for $\Xi^-$ and
  $\overline{\Xi}^+$ could only be   estimated: 
  we used the factors measured\cite{progress.report} for proton and
  antiproton:  $\alpha_{\Xi^-} =
  \alpha_{\overline{p}} = 1.3$ and
  $\alpha_{\overline{\Xi}^+} = \alpha_{p} = 0.7$. The enhancements as defined in
  equations   \ref{eq:enhancement-pA-iso} and   \ref{eq:enhancement-AA-iso}
  are shown in the panels (b) and (d) of figure \ref{fig:enhancement}
  for $\Xi^-$ and $\overline{\Xi}^+$, respectively. After taking into
  account the isospin effects the enhancement factors for $\Xi^-$ and
  $\overline{\Xi}^+$ approached each other, and p-Pb and Pb-Pb results
  are still of the same order.

\subsection{$\Omega$ Production in p-p Collisions}

\begin{figure}
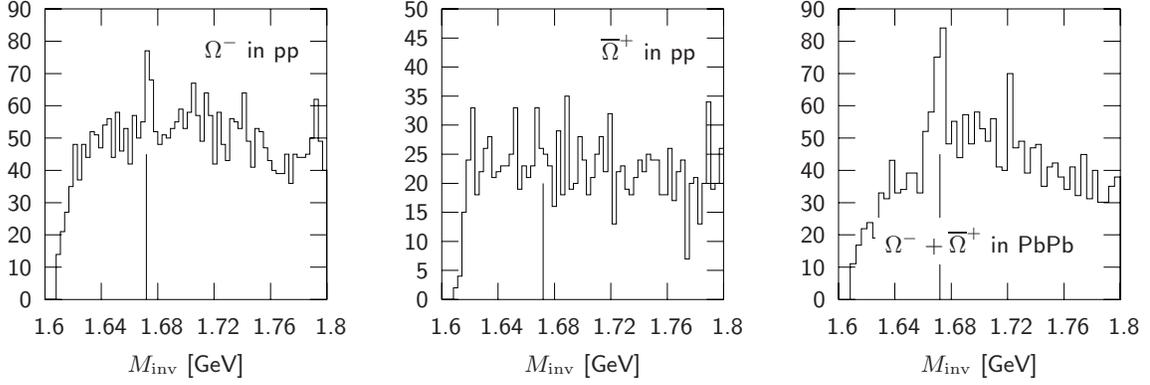

{\centerline{
\footnotesize\sf
\input{omega-pp}\hss
\input{aomega-pp}\hss
\input{omega-PbPb}
}}
\caption{Omega mass spectra in p-p and Pb-Pb collisions}
\label{fig:omega}
\end{figure}

$\Omega^-$ signals have been observed in p-p collisions for the first
time. Panels (a) and (b) of figure \ref{fig:omega} show the $\Omega^-$
and $\overline{\Omega}^+$ mass spectra in p-p collisions; panel (c)
shows the $\Omega^- + \overline{\Omega}^+$ mass spectrum in Pb-Pb
collisions. From the absence of the $\overline{\Omega}^+$ signal in
p-p events we estimate the $\overline{\Omega}^+ / \Omega^-$ ratio to
be less than 0.5 with a confidence level of 95\%. 

\section{Conclusions}

Results obtained from p-p and n-p collisions indicate a sizeable
isospin-dependence of the $K/\pi$ ratio. Using these data to eliminate
isospin effects from Pb-Pb results the $K/\pi$ ratio in Pb-Pb and p-Pb
collisions in the forward hemisphere agree within errors. 

The final $\Xi$ and feeddown-corrected $\Lambda$ results in p-p and
Pb-Pb collisions at the NA49 experiment are obtained. The
$\Xi$-enhancements at midrapidity in p-A and A-A collisions with
respect to p-p collisions are of the same order. 
An increased statistics should make the measurement of
$\Lambda$ and $\Xi$ particles in n-p collisions possible in the near
future. In the present study the isospin effect on $\Xi$ production in p-Pb and
Pb-Pb collisions was estimated using the factors determined for
protons and antiprotons from p-p and n-p collisions. 
 Isospin-corrected enhancement factors of
$\Xi^-$ and $\overline{\Xi}^+$ approach each other. 

An $\Omega^-$ signal has been observed in p-p collisions for the first
time. The $\overline{\Omega}^+/\Omega^-$ ratio is estimated to be less
than 0.5 with a confidence level of 95\%. $\Omega$ production has also
been observed in Pb-Pb collisions in NA49.

\section*{Acknowledgements}

This work was partially supported by the Hungarian Scientific Research
Fund (OTKA) under the contracts F034707 and T0032293.

\end{document}